\documentclass[12pt]{article}
\usepackage[dvips]{color}
\usepackage{amsmath}
\usepackage{graphicx}
\usepackage{subfigure}
\usepackage{epsfig}
\usepackage{amsmath}
\usepackage{cite}
\usepackage{color}
\usepackage{subfigure}
\usepackage{multirow}

\begin{document}
\begin{center}
\Large{\bf The Role of Topological Photon Spheres in Constraining the Parameters of Black Holes }\\
 \small \vspace{0.5cm}
 {\bf Jafar Sadeghi $^{\star}$\footnote {Email:~~~pouriya@ipm.ir}}, \quad
 {\bf Mohammad Ali S. Afshar $^{\star}$\footnote {Email:~~~m.a.s.afshar@gmail.com}}\\
\vspace{0.5cm}$^{\star}${Department of Physics, Faculty of Basic
Sciences,\\
University of Mazandaran
P. O. Box 47416-95447, Babolsar, Iran}\\
\small \vspace{0.5cm}
\end{center}
\begin{abstract}
In this paper, we investigate the topological photon sphere from two distinct perspectives. In the first view, we examine the existence and characteristics of topological photon(anti-photon)spheres for black holes with different structures, such as Einstein-Young-Mills non-minimal, AdS black holes surrounded by Chaplygin-like dark fluid, and Bardeen-like black holes in Einstein-Gauss-Bonnet gravity.
Furthermore, we delve into the deeper perspective of the necessity of photon spheres for super-compact gravitational structures such as black holes. By leveraging this necessity, we propose a classification of the parameter space of black hole models based on the existence and positioning of photon spheres. This approach enables the determination of parameter ranges that delineate whether a solution represents a black hole or a naked singularity.\\ In essence, the paper illustrates the utility of the photon sphere as a notable test for establishing the permissible and non-permissible parameter ranges within specific theories of black hole solutions.
\end{abstract}
\tableofcontents
\section{Introduction}
The fundamental concept of a black hole, which initially emerged as a theoretical construct and perhaps to some extent to lend practical credibility to the complex concepts of general relativity, has significantly evolved to this day. Observations today clearly indicate that they transcend mere abstract models or purely mathematical structures. The researchers actively expanded their understanding of general relativity instead of passively waiting for experimental confirmation.  They formulated various theoretical amendments in the gravitational equations and integrated new concepts of quantum gravity and string theory into this framework, which these efforts led to the proposal of various models for the structures of black holes, that each possessing unique structures and distinct parameters. One of our main objectives in this article is to examine the impact of these parameters and their permissible range, in several different models, on the topological photon spheres. As we know, the allowed range for these parameters is affected by the special preconditions that we impose on the model, which makes the range of their changes in different black holes not the same.
For instance, one of these restrictive general preconditions is the Weak Cosmic Censorship (WCC) conjecture, which states that singularities that appear when solving Einstein's equations must be hidden behind the event horizon and no singularity should be observable from the infinite future. In other words, in a singularity-free universe, as demonstrated by the theory of general relativity, the evolution of the universe is deterministic and predictable. A closer look at the above statement reveals that this situation is somewhat more indicative of a computational limitation than a scientific reality. In fact, our inability to access a stronger geometry and mechanics that can explain the complexities behind the event horizon has led us to discard anything that cannot be justified by our current understanding. However, over time, facing various conditions and the growth of knowledge has somewhat undermined the credibility of this condition because today, the potential emergence of points that we call "naked singularities" has become apparent in specific scenarios \cite{1,2,3}. In the past, it was always tried to avoid these singularities as much as possible due to the loss of space-time geometry, but today there are several studies on the continuous gravitational collapse of the inhomogeneous matter cloud that illustrates that space-time singularities developed during gravitational collapse can be seen by an outside observer and the visibility of space-time singularity depends upon the initial conditions of the collapsing matter\cite{4,5,6,7,8,9,10,11,12,13,14,15,16,17,18,19}.
This multiplicity of studies shows that, unlike previous studies in which only the permitted areas were considered, one should be sensitive to the classification of other areas because these areas may also be involved in the results under certain conditions, so their domain must be defined.
Given these interpretations, we must be able to categorize the range of our model parameters, i.e., know in what range the metric background of our black hole solution remains untouched, or it exhibits the behavior of a naked singularity, or the model completely falls apart. Distinguishing these domains and developing a relatively precise method to accomplish this task may be challenging. Nevertheless, the clichéd and fractal nature of the universe to some extent holds promise.
The fact that a structure must adhere to an almost constant pattern to have a black hole-like form has led us to the idea that perhaps, by using the method of inverse motion, it may be possible to identify experiments and approaches to determine the range of parameters. The method we propose for this purpose and finding the permissible range for parameters is the use of the concept of a topological photon sphere.
In classical physics, the emergence of rotational motion due to gravitational forces is a normal and obvious occurrence. The analogy of this classical pattern in cosmological models, as well as observational evidence, indicates that the existence of a photon ring is a necessity for the existence of a black hole. For example, in an observational experiment, researchers observed a dark shadow resulting from the bending of light by the massive black hole at the center of the giant elliptical galaxy M87, explicitly confirming the existence of photon rings around the central compact body \cite{20}. Of course, this is not the only example, and more such cases can be cited \cite{20,21,22,23,24,25}.  
From a theoretical perspective, the necessity of a photon ring has been examined in various models of gravitational potential in numerous articles \cite{26,27,28,29,30,31}, to the extent that today, the existence of photon rings for different black hole models can be somewhat assured.
So that today we can be sure of the existence of photon rings for black holes with high accuracy.\\ To calculate the photon sphere, you can follow different paths, for example, you can use the Hamiltonian, effective potential, and Kiling symmetries, or you can go for the topological method. In this article, we used the topological method. To achieve this method Cardoso et al. stated that" the nonlinear instability of ultra- compact stars would provide a strong argument in favor of the "black hole hypothesis," once electromagnetic or gravitational-wave observations confirm the existence of light rings", the objects with light rings are black holes \cite{32}.  In addition, Cunha and his colleagues showed that the standard black holes with spherical  symmetry, have  planar circular photon orbits, so that the stable type causes instability in spacetime and its unstable type can help us to determine the shadows of the black hole \cite{33}. Then, they proved a theorem that further strengthened this concept, a theorem that stated: "axisymmetric, stationary solutions of the Einstein field equations formed from classical gravitational collapse of matter obeying the null energy condition, that are everywhere smooth and ultra-compact (i.e., they have a light ring) must have at least two light rings, and one of them is stable. It has been argued that stable light rings generally lead to nonlinear spacetime instabilities\cite{34}. In this context, the work of Ghosh and Sarkar can also be mentioned \cite{35}. Expanding this hypothesis to four-dimensional black holes and testing its validity was the next step of this team, which showed: 
"A stationary, axisymmetric and asymptotically flat black hole space-time in 1+3 dimensions, with a non-extremal, topologically spherical and Killing horizon admits, at least, one standard light ring outside the horizon for each rotation sense" \cite{36}. At last, Shao-Wen Wei utilized this concept and Duan mapping, extending the discussion of the photon ring to the photon sphere with a topological approach"It is worth noting that there exists at least one standard photon sphere outside the black hole not only in asymptotically flat space-time but also in asymptotically Ads and dS space-time" \cite{37}.
Collectively, the comprehensive information presented above unequivocally confirms the presence of photon rings and spheres within black hole structures.
This essential feature of black hole structures motivated us to explore whether we can determine the permissible range of the specific parameters of each black hole with respect to the photon sphere. To achieve this goal and obtain the equations that define the permissible ranges for each parameter, we employed the method suggested by Wei to classify black holes based on their total topological charge. By examining the required conditions for the existence of a photon sphere for different black holes, we derived intervals for the specific parameter of each black hole, such that in these intervals:\\
1. The black hole possesses a standard unstable photon sphere, meaning that it exhibits the behavior of a normal black hole.\\
2. It approaches a naked singularity that has its own distinctive behaviors.\\
3. The absence of any photon structures, and thus the violation of the condition for the existence of a photon sphere, renders these intervals as prohibited for the structure to retain its classification as a black hole.\\\\
In general, this article can be examined from two separate perspectives, firstly, based on the topological method, we have examined the photon sphere for three separate and structurally different black hole samples.Secondly, we tried to obtain the allowed ranges for the parameters of each model by using the necessary conditions for the formation of the photon sphere.\\All above information give us motivation to arrange the paper as follows. In section 2 we briefly review the mathematical and physical foundations of the work in such a way that firstly, the effective potential and states the circular null geodesic conditions is introduced and then goes to the Duan’s topological mapping and at the end of this section, in an important conclusion, we recount the general way of working. In section 3,4,5 we will test the obtained technique on different black holes, and we will determine the allowed ranges of parameters with respect to different black holes. Finally, in section 6 we summarized our conclusions.

\section{ Standard and Topological Methods for Investigating the Photon Sphere }
Let's start with some commonly used definitions. The event horizon, a null one-sided hypersurface that acts as a causal boundary, ensuring that the weak cosmic censorship (WCC) holds. Just opposite is the naked singularity, as a hypothetical gravitational singularity without an event horizon. The photon sphere is a null ring that is the lower bound for any stable orbit, showing the extreme bending of light rays in a very strong gravity. It has two types: unstable, where small perturbations make the photons either escape or fall into the black hole, and stable, where the opposite happens. The unstable type is useful for determining the black hole shadows, while the stable type causes spacetime instability.\\
It is necessary to note that in this article we only consider and examine structures that have spherical symmetry, that is, models in the following form,
\begin{equation}\label{(1)}
\mathit{ds}^{2}=-\mathit{dt}^{2} f \! \left(r \right)+\frac{\mathit{dr}^{2}}{g \! \left(r \right)}+\left(d\theta^{2}+d\varphi^{2}
\sin \! \left(\theta \right)^{2}\right) h \! \left(r \right).
\end{equation}
In short, In the standard method to find the photon sphere, we go to the integral of the action and based on its lagrangian, we extract the Hamiltonian in the form of four-momentum, and then by applying the condition of having zero geodesics, we will have a relation in the following form,
\begin{equation}\label{(2)}
\Pi=\frac{p^{\nu} p^{\mu} \textit{g}_{\mu ,\nu}}{2}=0.
\end{equation}
Now, according to the obtained Hamiltonian form, we try to obtain the effective potential, which usually has the following standard form, 
\begin{equation}\label{3}
\begin{split}
V_{eff}=g(r)\bigg(\frac{L^2}{h(r)}-\frac{E^2}{f(r)}\bigg),
\end{split}
\end{equation}
where $E$ and $L$ represent the photon’s energy and the total angular momentum, respectively.
The radial component of the null geodesic equations will be as \cite{37},
\begin{equation}\label{4}
\begin{split}
\dot{r}^{2}+V_{eff}=0.
\end{split}
\end{equation}
A circular null geodesic occurs at an extremum of the effective potential $V_{eff}$(r), which is given by,
 \begin{equation}\label{5}
\begin{split}
V_{eff}=0,\hspace{1cm}\partial_{r}V_{eff}=0.
\end{split}
\end{equation}
These local extrema of the effective potential  are equivalent to unstable and stable circular null geodesics, will correspond to a photon sphere and an anti-photon sphere.
By Considering the equations(5) at the same time, we have  following expression,
\begin{equation}\label{6}
\begin{split}
\bigg(\frac{f(r)}{h(r)}\bigg)^{'}_{r=r_{ps}}=0,
\end{split}
\end{equation}
where prime is the derivative with respect to $r.$
By rewriting(6), we will have,
\begin{equation}\label{7}
\begin{split}
f(r)h(r)'-f(r)'h(r)=0.
\end{split}
\end{equation}
The first term at the horizon will be disappear, but this case the second term usually remains non-zero, it means that $r_{ps}$ and $r_{h}$ do not coincide\cite{37}.\\
In our assessment, the standard approach commonly employed to investigate the photon sphere possesses both merits and drawbacks. On one hand, this method is not contingent on the symmetry of the metric function, rendering it applicable to any metric form. However, in highly intricate structural models, it may be challenging to straightforwardly ascertain the Hamiltonian function, or even if it is determined, extracting its effective potential may prove unfeasible.
Furthermore, the differentiation between stable and unstable photon spheres necessitates extensive computations. This implies that relying on the roots of the metric function can only delineate the permissible regions for a black hole-like structure. Essentially, we are only able to discern the range of parameters with which unstable photon spheres can be precisely characterized. Perhaps this is the motivation that we needed to lead us to the topological method.\\ 
Yishi Duane, in $1984$, in examining the inner structure of the conserved topological currents in SU(2) theory, introduced the topological flow related to a point system like particles, which a foundation to discuss other kinds of topological currents\cite{38}.
In the first step, we consider a general vector field as  $\phi$  which can be decomposed into two components, $\phi^r$ and $\phi^\theta$,
\begin{equation}\label{8}
\phi=(\phi^{r}, \phi^{\theta}),
\end{equation}
also here, we can  rewrite the vector as $\phi=||\phi||e^{i\Theta}$, where $||\phi||=\sqrt{\phi^a\phi^a}$, or $\phi = \phi^r + i\phi^\theta$.
Based on this, the normalized vector is defined as,
 \begin{equation}\label{9}
n^a=\frac{\phi^a}{||\phi||},
\end{equation}
where $a=1,2$  and  $(\phi^1=\phi^r)$ , $(\phi^2=\phi^\theta)$.
Now we introduce our antisymmetric superpotential as follows\cite{37,38},
 \begin{center}
 $V^{\mu\nu}=\frac{1}{2\pi} \epsilon^{\mu\nu\rho} \epsilon_{ab}n^a\partial_\rho n^b,\hspace{0.3cm}\mu,\nu,\rho=0,1,2$,
\end{center}
and the topological current will be as,
\begin{center}
 $j^{\mu}=\partial_{\nu}V^{\mu\nu}=\frac{1}{2\pi}\epsilon^{\mu\nu\rho}\epsilon_{ab}\partial_{\nu}n^a \partial_\rho n^b$.
\end{center}
Based on this, Noether's current and charges at the given $\Omega$ will be,
\begin{center}
 $\partial_\nu j^\nu=0$,
 \end{center}
and
\begin{equation}\label{10}
Q=\int_{\Omega}j^0d^{2}x,
\end{equation}
where $j^0$ is the charge density.
By replacing $\phi$ instead of $n$ and using the Jacobi tensor, we will arrive at,
\begin{equation}\label{(11)}
j^{\mu}=\frac{J^{\mu} \! (\frac{\phi}{x}) \ln \! \left({||\phi||}\right) \Delta_{\phi^{a}}}{2 \pi}.
\end{equation}
By using  two-dimensional Laplacian Green function in $\phi-mapping$ space, we have,
\begin{equation}\label{(12)}
\ln \! \left({||\phi||}\right) \Delta_{\phi^{a}}=2 \delta \! \left(\phi \right) \pi,
\end{equation}
and the topological current  will be,
\begin{equation}\label{(13)}
j^{\mu}=J^{\mu} \! \left(X \right) \delta^{2} \! \left(\phi \right).
\end{equation}
From the properties of $\delta$, it is clear that $j^{\mu}$ is non-zero only at the zero points of $\phi^{a}$, and this is exactly what we need to continue the discussion.
Finally, by using the above relations and inserting them in relation (10), the topological charge become,
\begin{equation}\label{14}
Q=\int_{\Omega}J^{0} \! \left(X \right) \delta^{2} \! \left(\phi \right)d^{2}x,
\end{equation}
once again and this time around the topological charge Q, according to the characteristic of the $\delta$ function, it can be said that the charge is non-zero only at the zero point of $\phi$.\\
Now, according to the above statements, we will go to the definition of the potential for the topological model, according to \cite{37} we will have
\begin{equation}\label{15}
\begin{split}
H(r, \theta)=\sqrt{\frac{-g_{tt}}{g_{\varphi\varphi}}}=\frac{1}{\sin\theta}\bigg(\frac{f(r)}{h(r)}\bigg)^{1/2},
\end{split}
\end{equation}
the discussion of potential will allow us to search for the radius of our photon sphere at,
\begin{center}
  $\partial_{r}H=0$,
 \end{center}
 so, we can use a vector field $\phi=(\phi^r,\phi^\theta)$,
\begin{equation}\label{16}
\begin{split}
&\phi^r=\frac{\partial_rH}{\sqrt{g_{rr}}}=\sqrt{g(r)}\partial_{r}H,\\
&\phi^\theta=\frac{\partial_\theta H}{\sqrt{g_{\theta\theta}}}=\frac{\partial_\theta H}{\sqrt{h(r)}}.
\end{split}
\end{equation}
With the above definition $\phi$, and recalling the relationships of the Duan’s section, we now define the current and charge (10) for this new potential.\\
Also, considering relation (13) and according to the characteristics of the $\delta$ function,one can say that the charges will be non-zero only at the points where $\phi$ is zero. That is, right where the photon sphere is placed, so we can attribute a topological charge Q to each photon sphere. According to  equation (13), when we take $\Omega$  as a manifold which  covers one zero point, in that case the investigations show that the charge Q is exactly equal to the number of winding\cite{37}.
Let $C_i$ be a closed curve that is smooth, positive oriented, and encloses only the i-th zero point of $\phi$, while the other zero points are outside of it. So the winding number can be calculated by the following formula,
\begin{equation}\label{17}
\omega_i=\frac{1}{2\pi}\oint_{C_{i}}d\Lambda,
\end{equation}
where $\Lambda$ is
\begin{equation}\label{18}
\Lambda=\frac{\phi^2}{\phi^1},
\end{equation}
then the total charge will be,
\begin{equation}\label{19}
Q=\sum_{i}\omega_{i}.
\end{equation}
Finally, when the closed curve includes a zero point, the investigation of corresponding note show that the charge Q is exactly equal to the winding number.  In that case,  if it covers multiple zero points, Q will be sum of the winding number at each zero point  and if encloses no zero point then we must have zero charge.\\
Before the end of this section it is better to explain, In a general conclusion, how to calculate topological charges and total charges for studying each case.\\ 
When the contour includes a winding:\\
1.If the field lines diverge around the zero point, its topological  charge is +1 (purple loop in Fig 3a). According to the previous definitions, this state will correspond to the naked singularity.\\  
2. If the field lines converge around the zero point, its topological  charge is -1 (purple loop in Fig 2). This state will correspond to a black hole with unstable photon sphere.\\  
3.If the contour does not include a winding, its charge will be zero.Finally, the total charges for each shape are the algebraic sum of the charges, which can be  displayed by drawing a contour that includes all the zero points.\\ 
Based on the aforementioned findings, it can be affirmed that each black hole may harbor a photon sphere. In the same way, the topological current is only non-zero at the zero point of the vector field, which determines the photon sphere's location. Consequently, a single topological charge can be attributed to each photon sphere. Previous research indicates that in the complete exterior region, when the structure takes the form of a black hole, the total topological charge consistently amounts to -1\cite{37}. Conversely, for a naked singularity with a vanishing topological charge, the total topological charges in this region will be either zero or +1\cite{37,39}.\\
Therefore, in light of the above information, the region where the behavior of the black hole results in a total topological charge of -1 aligns with the permissible values and our preferred parameter distance in the metric, corresponding to the inherent behavior of a black hole. 
Conversely, the region where the equations' behavior yields a total topological charge of zero or 1 will denote an interval for the parameters steering the system toward the naked singularity.
In conclusion, it can be inferred that beyond these two ranges, and due to the contravention of the photon sphere condition in these areas, these regions will likely be deemed illicit areas for the parameters.\\
$ R_{PLPS} $: \textit{According to the allowed range of parameters to preserve the structure of the black hole, the minimum or maximum possible radius for the appearance of an unstable photon sphere is called the radius $ R_{PLPS} $ (Possible Radius Limit for the Photon Sphere).}

\section{ Einstein-Young-Mills non-minimal regular black hole }
Regular non-minimal Einstein Yang-Mills (R-nm-EYM) black holes are a class of black hole solutions in a modified gravity theory that couples the curvature of spacetime to non-Abelian gauge fields. These black holes have two distinctive features: they are regular, meaning that they do not have a singularity at the center, and they have a magnetic charge, which is related to the Yang-Mills field strength.The EYM theory is a generalization of the Einstein-Maxwell theory, which couples the curvature of spacetime to the electromagnetic field. This theory replaces the electromagnetic field with a non-Abelian gauge field, which is a type of field that has internal symmetries and self-interactions. The EYM theory also introduces a non-minimal coupling term, which is a term that involves the product of the curvature tensor and the gauge field strength tensor. This term modifies the gravitational and gauge field equations, and leads to new types of solutions.The EYM theory and its black hole solutions are interesting for several reasons. They provide a way to study the effects of non-Abelian gauge fields and non-minimal couplings on the geometry and physics of black holes. They also offer a possibility to avoid the singularity problem, which is one of the major challenges in classical and quantum gravity. Moreover, they have potential applications to astrophysics and cosmology, as they could describe some of the observed phenomena related to black holes, such as gravitational lensing, gravitational waves, and black hole shadows\cite{40,41,42,43,44,45,46}.\\
the metric function for the 4D spherically symmetric R-nm-EYM black hole is given by \cite{40},
\begin{equation}\label{20}
f =1+\frac{r^{4} \left(-\frac{2 m}{r}+\frac{q^{2}}{r^{2}}+\frac{r^{2}}{l^{2}}\right)}{r^{4}+2 q^{2} \xi},
\end{equation}
where  $\xi$ is the nonminimal parameter  of the theory, $l$ is related to the cosmological constant( $l^2 =-\frac{3}{\Lambda}$), $q$ is the magnetic charge of the Wu-Yang gauge field,  and $m$ is the asymptotic mass. Since our main purpose is to study the allowed range for specific parameters of the theory, for this reason we value the usual parameters such as mass, AdS radius and charge, where these selected values are $m=1,l=1,q=0.5$.\\
Now, if we solve the metric function (20) based on the assumed values to find the event horizon in terms of $\xi$, we will find that our metric function always has a solution for $\xi\leq 0.41914 $ , fig(1a), and in better words, the structure can have the form of a black hole. But, for values greater than 0.41914, fig(1b), the metric function is always unanswered, which means that the structure has no event horizon, and as a result, it will be in the form of a naked singularity.
\begin{figure}[h!]
 \begin{center}
 \subfigure[]{
 \includegraphics[height=5.5cm,width=6cm]{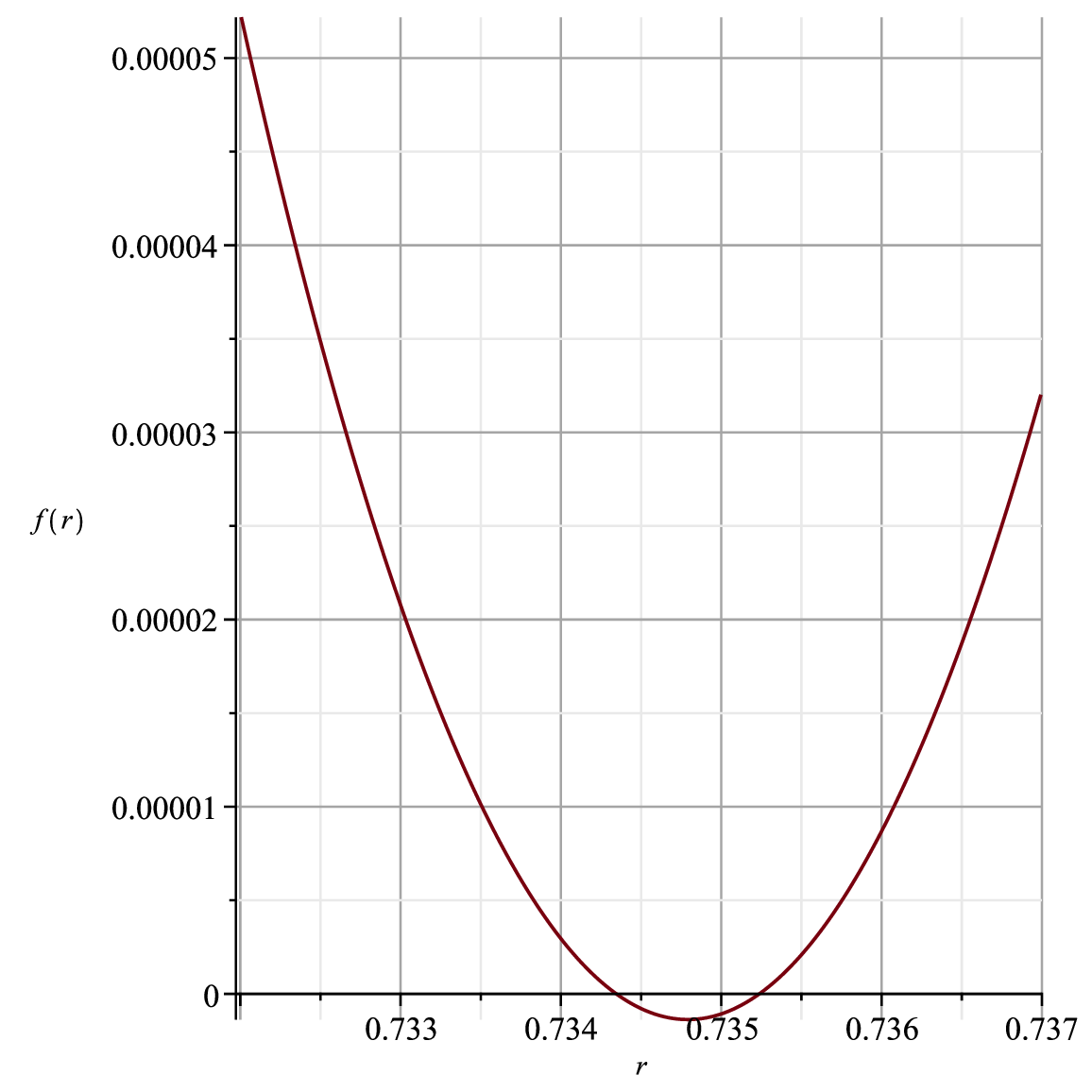}
 \label{1a}}
 \subfigure[]{
 \includegraphics[height=5.5cm,width=6cm]{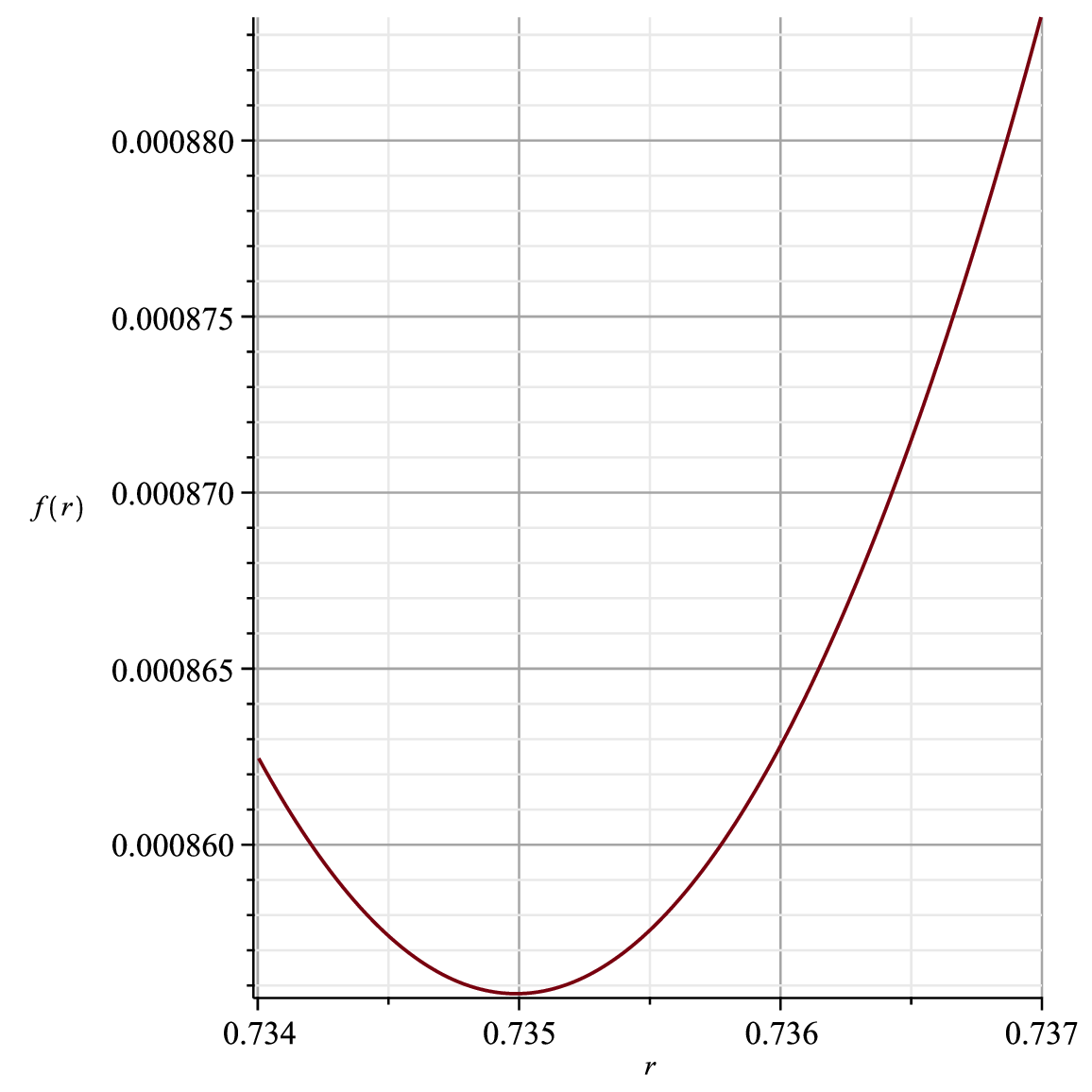}
 \label{1b}}
   \caption{\small{Plot 1(a) represents $f(r)$ vs. r with $\xi=0.41914 $  , Plot 1(b) shows the behavior $f(r)$ vs. r with $\xi=0.42 $  }}
 \label{1}
\end{center}
 \end{figure}
Now, to deal with the topological view, according to the parametric values mentioned above and since we have the following conditions for this model
\begin{equation}\label{(21)}
f \! \left(r \right)=g \! \left(r \right),
\end{equation}
\begin{equation}\label{(22)}
h \! \left(r \right)=r,
\end{equation}
from equations (15) and (16), we will have,
\begin{equation}\label{23}
H =\frac{\sqrt{1+\frac{r^{4} \left(-\frac{2}{r}+\frac{ 0.25}{r^{2}}+r^{2}\right)}{r^{4}+ 0.50 \xi}}}{\sin \! \left(\theta \right) r},
\end{equation}
\begin{equation}\label{24}
\phi^{r}=\frac{-r^{8}+ 3.0 r^{7}+\left(\xi - 0.5\right) r^{6}- 1.0 r^{4} \xi - 0.5 r^{3} \xi - 0.25 \xi^{2}}{r^{2} \sin \! \left(\theta \right) \left(r^{4}+ 0.5 \xi \right)^{2}},
\end{equation}
\begin{equation}\label{25}
\phi^{\theta}=-\frac{\sqrt{\frac{r^{4}+ 0.5 \xi - 2.0 r^{3}+ 0.25 r^{2}+r^{6}}{r^{4}+ 0.5 \xi}}\, \cos \! \left(\theta \right)}{\sin \! \left(\theta \right)^{2} r^{2}}.
\end{equation}
\subsection{ photon sphere ranges}
As we can see in fig(2), in the allowed areas for $\xi$,($ 0<\xi\leq 041914 $) the structure has a total topological charge of -1 and the black hole shows a normal behavior by showing unstable photon sphere. What can be pointed out here and is important is that, as shown in Table (1), to have a black hole form with respect to the main pre-selected parameters, is the restriction that we will have for the choice of $\xi$. In addition, with this method, the radius $ R_{PLPS} $ can also be calculated based on parametric constraints.
\begin{figure}[h!]
 \begin{center}
 \includegraphics[height=7.5cm,width=6.8cm]{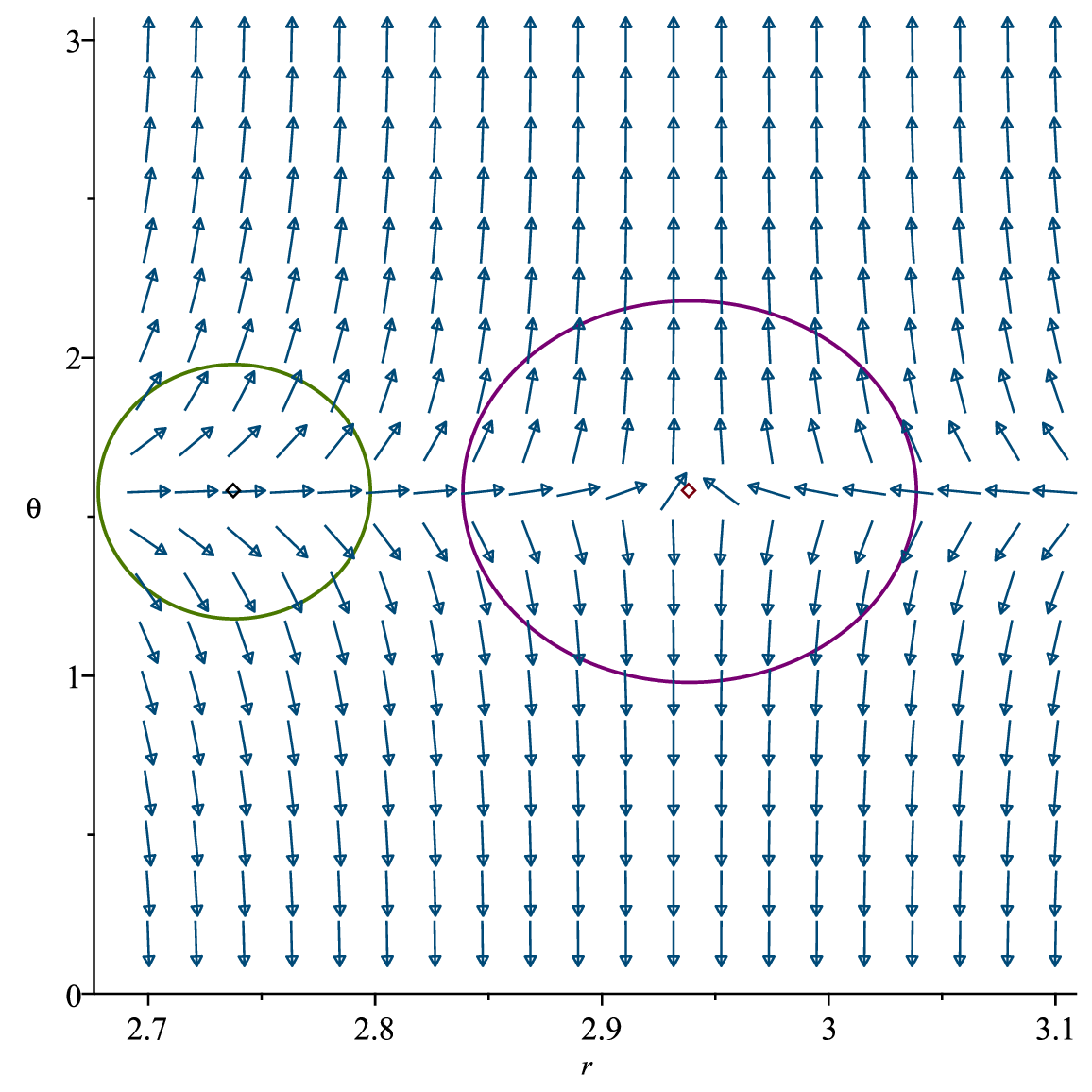}
 \caption{\small{the blue arrows represent the vector field n on a portion of the$ (r , \theta)$ plane, The photon sphere is located at $ (r,\theta)=(2.938684304,1.57)$  with respect to $(q=0.5,m=1,\xi=0.37,l=1)$}}
 \label{2}
\end{center}
\end{figure}
\subsection{naked singularity ranges}
From a mathematical point of view, using equations (24) and (25) to find the topological charges, we find that unlike the metric function which assumes that for all values of $ \xi \leq 0.41914 $ the structure has an event horizon(that is, it can still behave like a black hole), but Our calculations for the $\phi$ function show that in negative $ \xi$ regions, the structure only responds up to $ \xi = -4.3605 $, although this response is in the form of a naked singularity fig(3a) and after this range the response completely disappears. This is to some extent a confirmation and maybe a complement to the statement that was expressed in the \cite{40} article. Also, the calculations for the positive regions show that the structure for $\xi$'s greater than 0.41914 with a total topological charge of zero behaves in the form of a naked singularity (Fig 3b), which means that there is a limit for choosing positive z's as well.
\begin{figure}[h!]
 \begin{center}
 \subfigure[]{
 \includegraphics[height=5.5cm,width=6cm]{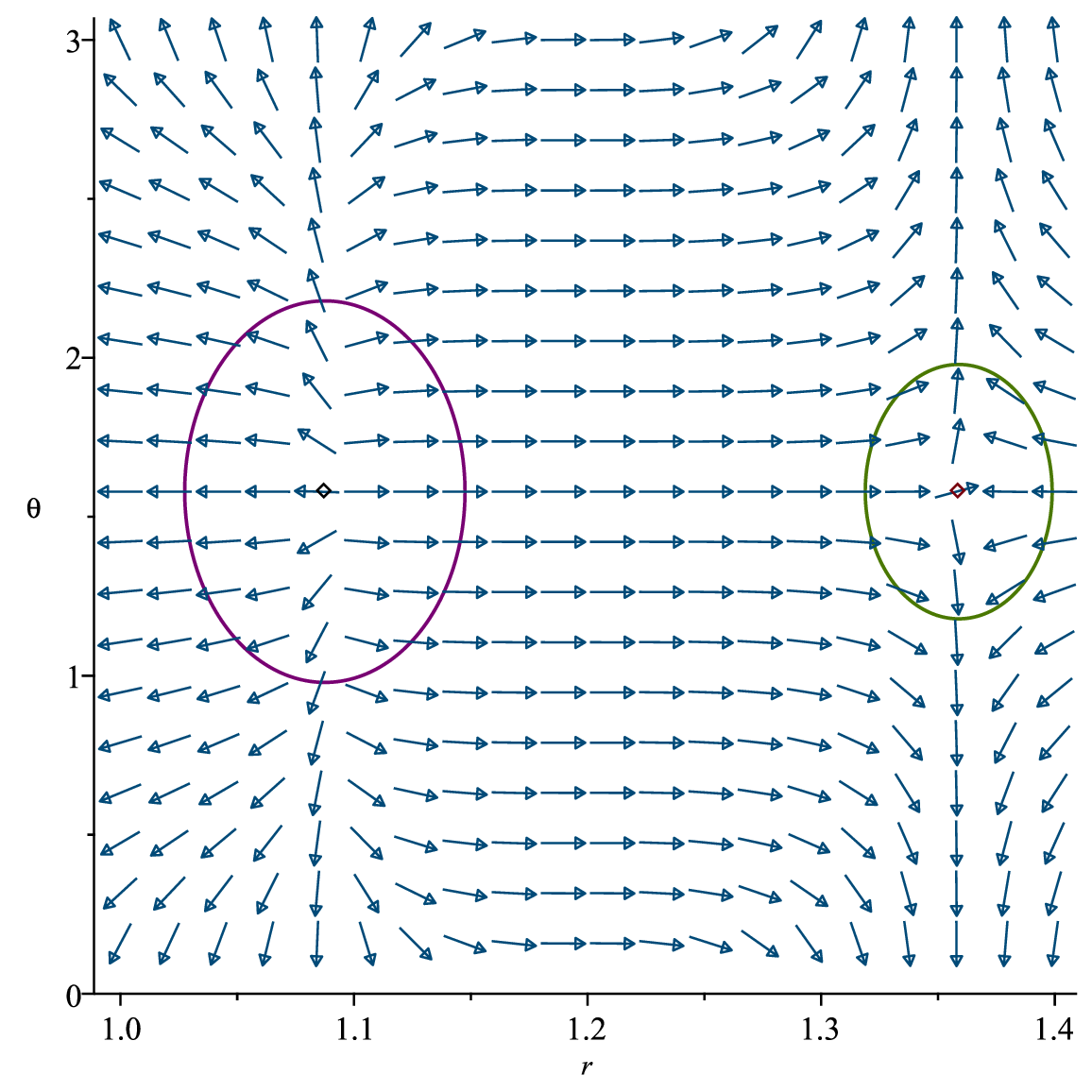}
 \label{3a}}
 \subfigure[]{
 \includegraphics[height=5.5cm,width=6cm]{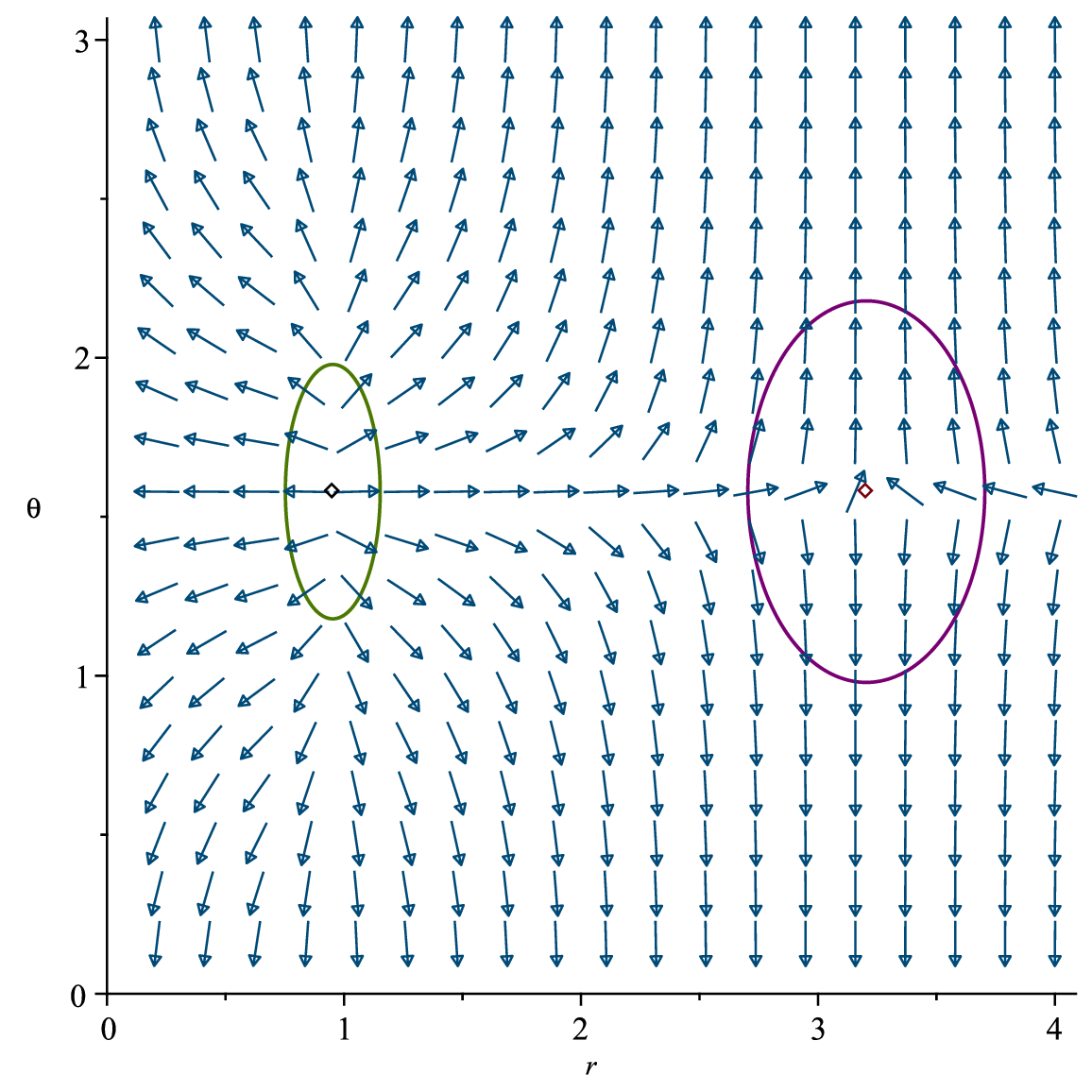}
 \label{3b}}
   \caption{\small{Plot 3(a) The photon spheres is located at $ (r,\theta)=(1.087510124,1.57)$ and $ (r,\theta)=(1.358858788,1.57)$ with respect to $(q=0.5,m=1,\xi=-4.1,l=1)$ , Plot 3(b) The photon spheres is located at $ (r,\theta)=(0.9525586040,1.57)$ and $ (r,\theta)=(3.203877596,1.57)$ with respect to $(q=0.5,m=1,\xi=1.3,l=1)$ }}
 \label{3}
\end{center}
 \end{figure}
The full results of this section are given in Table (1).
\begin{center}
\begin{table}[h!]
  \centering
\begin{tabular}{|p{3.2cm}|p{4cm}|p{4cm}|p{1.5cm}|p{2cm}|}
  \hline
  \centering{R-nm-EYM black holes} & \centering{Fix parameters} &\centering{Conditions} & \centering *TTC&\ $(R_{PLPS})$\\[3mm]
   \hline
  \centering{*Unauthorized area} & $q=0.5,m=1,l=1$ & \centering{$\xi<-4.3605$} & $ nothing$&\ $-$\\[3mm]
   \hline
  \centering{naked singularity} & $q=0.5,m=1,l=1$ & \centering{$-4.3605\leq\xi<0$} & $0$&\ $-$ \\[3mm]
   \hline
   \centering{unstable photon sphere} & $q=0.5,m=1,l=1$ & \centering{$0<\xi\leq0.41914$} & $ -1 $&\ $2.953579630$ \\[3mm]
   \hline
   \centering{naked singularity} & $q=0.5,m=1,l=1$ & \centering{$0.41914<\xi$} & $ 0 $&\ $-$ \\[3mm]
   \hline
      \end{tabular}
   \caption{*Unauthorized region: is the region where the roots of $\phi$ equations become negative or imaginary in this region.\\ TTC: *Total Topological Charge\\}\label{1}
\end{table}
 \end{center}
\section{ AdS black holes surrounded by dark fluid with Chaplygin-like
equation of state }
AdS black holes surrounded by dark fluid with Chaplygin-like equation of state are a class of black hole solutions in a modified gravity theory that includes a dark fluid as a cosmic background. The dark fluid is a hypothetical form of matter that could account for the accelerated expansion of the universe. The dark fluid has a Chaplygin-like equation of state, which means that its pressure is inversely proportional to its density. The Chaplygin-like equation of state can interpolate between different regimes, such as dust, radiation, cosmological constant, and phantom energy. The dark fluid can also mimic the effects of dark matter and dark energy, which are the two main components of the dark sector of the universe. The AdS black holes surrounded by dark fluid with Chaplygin-like equation of state are interesting for several reasons. They provide a way to study the effects of dark fluid and Chaplygin-like equation of state on the geometry and physics of black holes. They also offer a possibility to resolve the singularity problem and to describe a regular black hole. Moreover, they have potential applications to astrophysics and cosmology, as they could describe some of the observed phenomena related to black holes, such as gravitational lensing, gravitational waves, and black hole shadows.\cite{47,48,49}\\
the metric function for the 4D spherically symmetric Chaplygin-like Dark Fluid black holes is given by \cite{47},
\begin{equation}\label{26}
f =1-\frac{2 m}{r}+\frac{\left(\frac{q}{r^{3} \sqrt{b}}+\frac{q^{3}}{6 r^{9} b^{\frac{3}{2}}}\right) q}{3 r}-\frac{\sqrt{b +\frac{q^{2}}{r^{6}}}\, r^{2}}{3}+\frac{r^{2}}{l^{2}},
\end{equation}
where in this case, the Chaplygin-like Dark Fluid has a non-linear form as  $p = -b/\rho$ , that $\rho$  is energy density and ; b ; is a positive constant, $l$ is related to the cosmological constant( $l^2 =-\frac{3}{\Lambda}$),  $q>0$,  is a normalization factor that indicates the intensity of the Chaplygin-like Dark Fluid,  and $m$ denotes the mass of the black hole. Since our main purpose is to study the allowed range for specific parameters of the theory, for this reason we value the usual parameters such as mass, AdS radius , where these selected values are $m=1,l=1 $.\\
Now, if we solve the metric function (20) based on the assumed values to find the event horizon in terms of $b = 0.76$, we will find that our metric function always has a solution for$, 0<q \leq 1.02709 $, fig(4), and in better words, the structure can have the form of a black hole.But, for values greater than 1.02709, the metric function is always unanswered, which means that the structure has no event horizon, and as a result, it will be in the form of a naked singularity.
\begin{figure}[h!]
 \begin{center}
 \includegraphics[height=7.5cm,width=6.8cm]{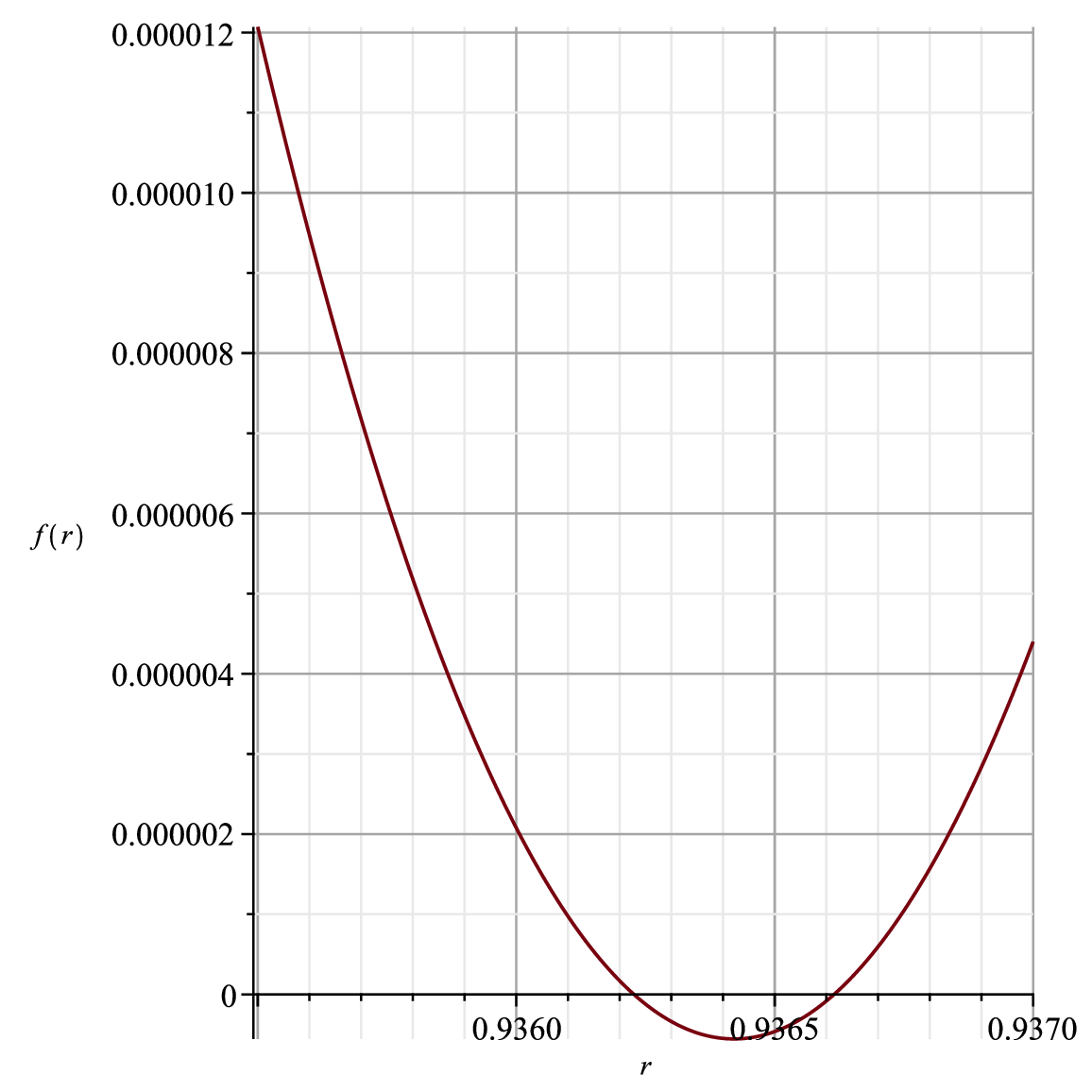}
 \caption{\small{this plot represents ($f(r)$ vs. r) with $b = 0.76$ and $q = 1.02709 $ where the event horizon located at $ r_+ = 0.9366148712$}}
 \label{2}
\end{center}
\end{figure}
Now, to deal with the topological view, according to the parametric values mentioned above and with respect to (21) and (22), for such black hole, the potential H and $\phi$ (15) , (16) after some calculations are,
\begin{equation}\label{27}
H =\frac{\sqrt{9-\frac{18}{r}+\frac{3 \left(\frac{q}{r^{3} \sqrt{b}}+\frac{q^{3}}{6 r^{9} b^{\frac{3}{2}}}\right) q}{r}-3 \sqrt{b +\frac{q^{2}}{r^{6}}}\, r^{2}+9 r^{2}}}{3 \sin \! \left(\theta \right) r},
\end{equation}
\begin{equation}\label{28}
\begin{split}
&\mathcal{A}=((r^{9} (r -3) b^{\frac{3}{2}}+b \,q^{2} r^{6}+\frac{q^{4}}{3}) \sqrt{\frac{b \,r^{6}+q^{2}}{r^{6}}}-\frac{q^{2} r^{6} b^{\frac{3}{2}}}{2}) \sqrt{3}\\
&\times\sqrt{9-\frac{18}{r}+\frac{3 (\frac{q}{r^{3} \sqrt{b}}+\frac{q^{3}}{6 r^{9} b^{\frac{3}{2}}}) q}{r}-3 \sqrt{b +\frac{q^{2}}{r^{6}}}\, r^{2}+9 r^{2}}\\
&\mathcal{B}=3 \sqrt{\frac{-\sqrt{\frac{b \,r^{6}+q^{2}}{r^{6}}}\, r^{12} b^{\frac{3}{2}}+(3 r^{12}+3 r^{10}-6 r^{9}) b^{\frac{3}{2}}+b \,q^{2} r^{6}+\frac{q^{4}}{6}}{b^{\frac{3}{2}} r^{10}}}\, \sqrt{\frac{b \,r^{6}+q^{2}}{r^{6}}}\, b^{\frac{3}{2}} r^{12} \sin \! (\theta )\\
\phi^{r}=\frac{\mathcal{A}}{\mathcal{B}},
\end{split}
\end{equation}
\begin{equation}\label{29}
\phi^{\theta}=-\frac{\sqrt{9-\frac{18}{r}+\frac{3 \left(\frac{q}{r^{3} \sqrt{b}}+\frac{q^{3}}{6 r^{9} b^{\frac{3}{2}}}\right) q}{r}-3 \sqrt{b +\frac{q^{2}}{r^{6}}}\, r^{2}+9 r^{2}}\, \cos \! \left(\theta \right)}{3 \sin \! \left(\theta \right)^{2} r^{2}}.
\end{equation}
Calculations for this structure show that at a certain(optional) $b = 0.76$, the black hole in the allowed range of $ 0<q \leq 1.02709 $ in fig (5a) shows a total topological charge of -1 and has an unstable photon sphere. The important point about this fig (5a) is that two topological charges are seen in it, and therefore the total charge must be zero.
What should be noted is that the radius in which one of the charges appears, the charge enclosed in the green ring, is smaller than the radius of the event horizon $r_+ = 0.9366148712$, that is, this charge is located behind the horizon and practically according to the concept of WCC it will not be included in the calculations. In the interval $, 1.02709<q \leq 3.6282 $, it tends to naked singularity and having two separate charges outside the horizon of fig (5b), it shows zero topological charge. Finally, upon reaching the $ q = 3.6282 $ limit, for $q \geq 3.6283 $ the structure lacks any solution. The results of this section are shown in Table 2. 
\begin{figure}[h!]
 \begin{center}
 \subfigure[]{
 \includegraphics[height=5.5cm,width=6cm]{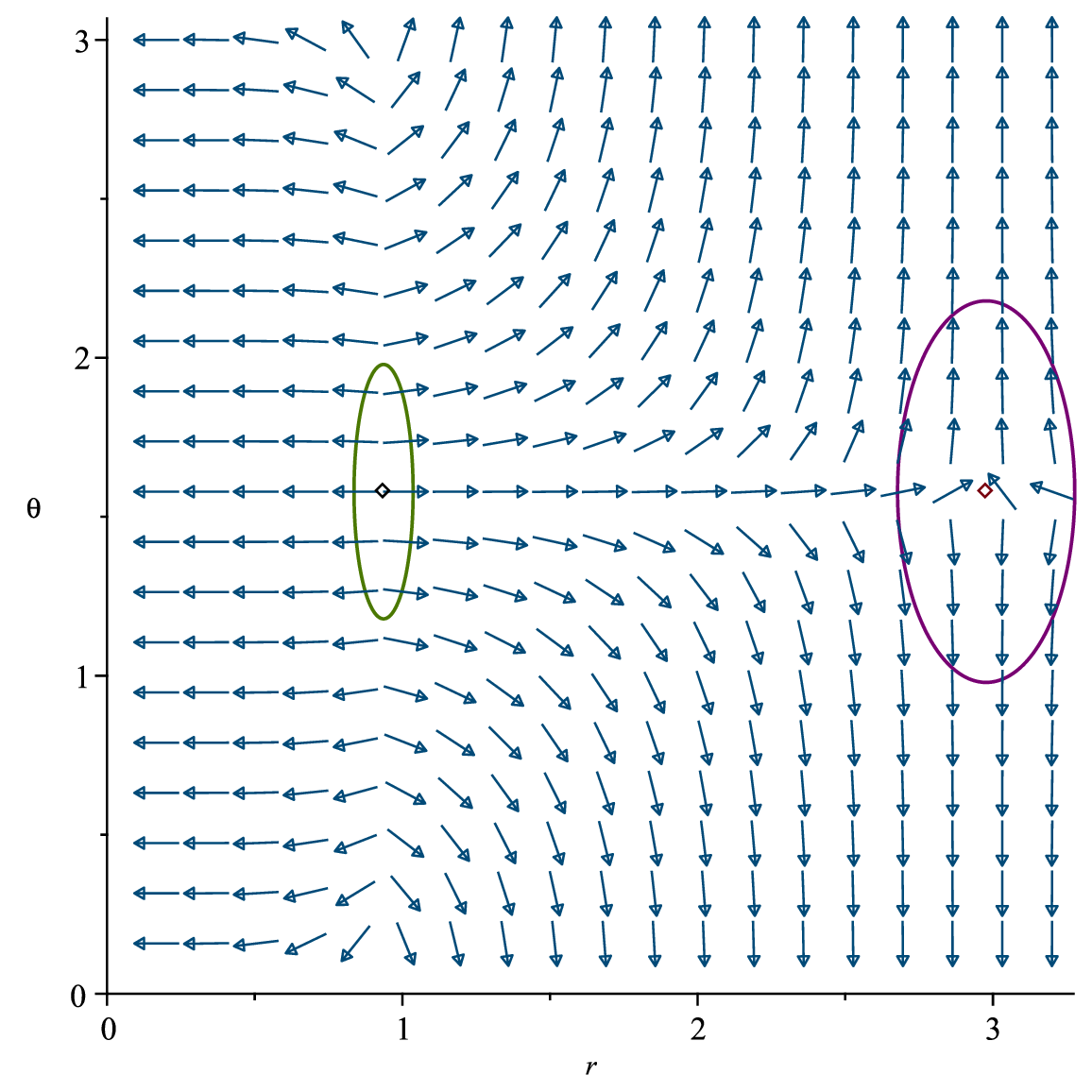}
 \label{5a}}
 \subfigure[]{
 \includegraphics[height=5.5cm,width=6cm]{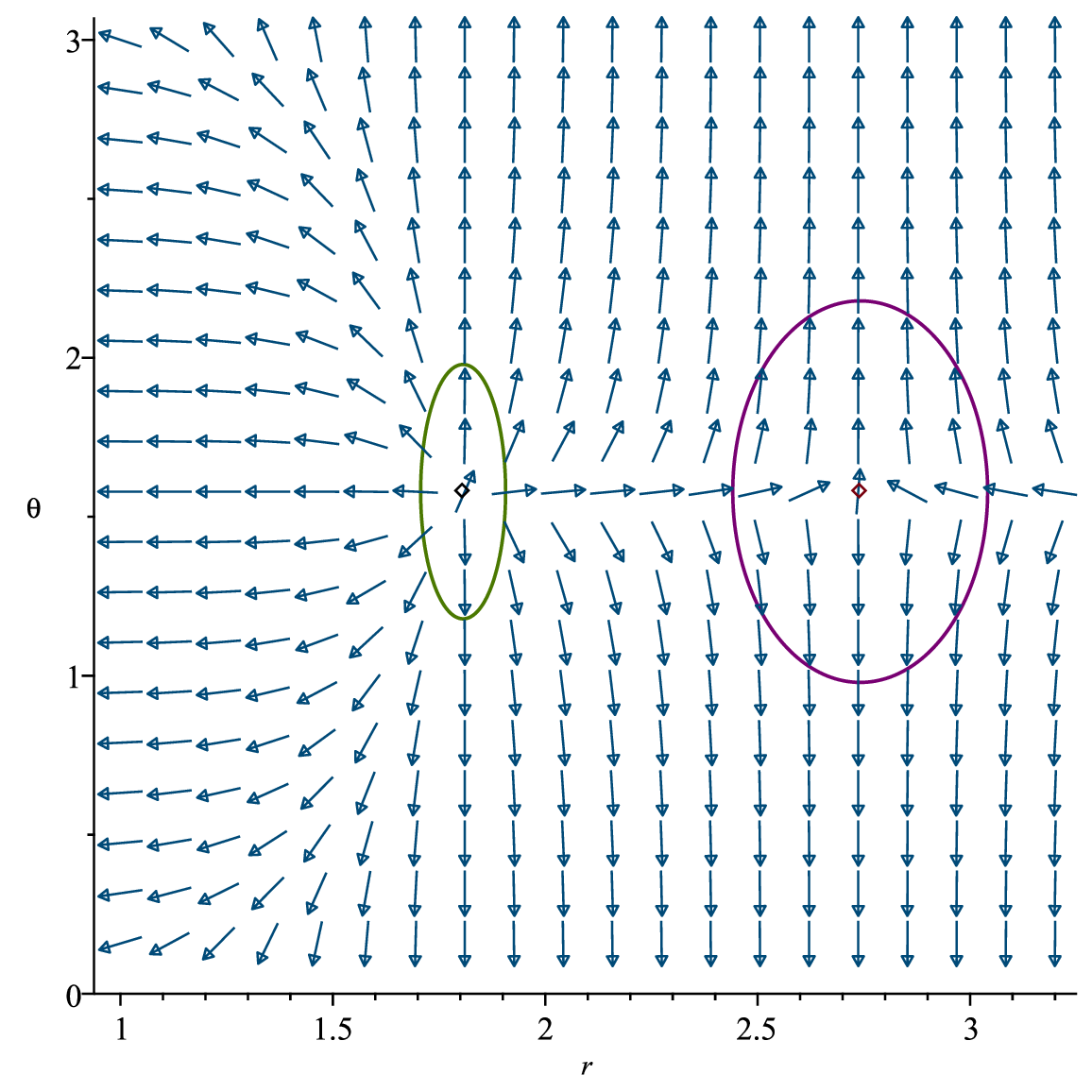}
 \label{5b}}
   \caption{\small{Plot 5(a) the photon spheres is located at $ (r,\theta)=(0.9364211602,1.57)$ and $ (r,\theta)=(2.977014940,1.57)$ with respect to $(b=0.76,m=1,q=1.02709,l=1)$ , Plot 5(b) the photon spheres is located at $ (r,\theta)=(1.806680305,1.57)$ and $ (r,\theta)=(2.741356787,1.57)$ with respect to $(b=0.76,m=1,q=3,l=1)$ }}
 \label{5}
\end{center}
 \end{figure}
\begin{center}
\begin{table}[h!]
  \centering
\begin{tabular}{|p{3.2cm}|p{4cm}|p{4.5cm}|p{1.5cm}|p{2cm}|}
  \hline
  \centering{Chaplygin-like Dark Fluid black holes}  & \centering{Fix parameters} &\centering{Conditions}&*TTC&\ $(R_{PLPS})$\\[3mm]
   \hline
  \centering{unstable photon sphere} & $b=0.76,m=1,l=1$ & \centering{$0<q \leq 1.02709$} &\centering{ $-1$}&\ $2.977014940$\\[3mm]
   \hline
  \centering{naked singularity} & $b=0.76,m=1,l=1$ & \centering{$1.02709<q \leq 3.6282$} &\centering{ $0$}&\ $-$ \\[3mm]
   \hline
   \centering{*Unauthorized area} & $b=0.76,m=1,l=1$ & \centering{$q \geq 3.6283 $} & $ nothing$&\ $-$\\[3mm]
   \hline
      \end{tabular}
   \caption{*Unauthorized region: is the region where the roots of $\phi$ equations become negative or imaginary in this region.\\ TTC: *Total Topological Charge\\}\label{1}
\end{table}
\end{center}
\section{ Bardeen-like regular black holes in Einstein-Gauss-Bonnet gravity }
The regularized  Einstein–Gauss–Bonnet gravity is a recent modification of the Einstein's general relativity that incorporates higher-order curvature terms in the action, namely the Gauss–Bonnet (GB) term, which is usually relevant only in higher dimensions. The regularization procedure, introduced by Glavan and Lin in 2020, allows the GB term to have a nontrivial effect in four dimensions by rescaling the GB coupling constant by the inverse of the space-time dimension and taking the limit as the dimension approaches four. This way, the field equations remain second-order and free of ghosts, but also acquire some novel features that differ from the standard Einstein gravity. Several authors have studied the properties of Bardeen black holes in the regularized  Einstein–Gauss–Bonnet gravity, such as their horizon structure, thermodynamics, stability, phase transitions, and strong gravitational lensing. Some of the main results are:\\
•  The GB parameter affects the minimum cutoff values of mass and charge for the existence of a black hole horizon, and also the size and shape of the horizons. Depending on the values of the parameters, there can be zero, one, or two horizons, corresponding to a naked singularity, an extremal black hole, or a regular black hole, respectively \cite{51,52}.\\
•  The black hole undergoes a second-order phase transition at a critical value of the horizon radius, where the temperature has a local maximum and the heat capacity diverges. The phase transition is similar to the Hawking–Page transition in AdS space, where the black hole becomes thermodynamically preferred over the pure de Sitter space below the critical temperature \cite{51}.\\
•  The black hole acts as a strong gravitational lens in the strong deflection limit, where the light rays can loop around the black hole multiple times before reaching the observer. The GB parameter and the NED parameter affect the observable of the strong gravitational lensing, such as the deflection angle, the position and magnification of the relativistic images, and the time delay between them \cite{53}.\\
We studied this structure separately with and without the cosmological constant in 4 and 5 dimensions.
\subsection{$5D:\Lambda=0 $}
the metric function for the 5D spherically symmetric Bardeen-like regular black holes in Einstein-Gauss-Bonnet gravity is given by \cite{50},
\begin{equation}\label{30}
f =1+\frac{r^{2} \left(1-\sqrt{1+\frac{8 \alpha  m}{\left(g^{2}+r^{2}\right)^{\frac{4}{3}}}}\right)}{2 \alpha},
\end{equation}
where m and g, are respectively mass and magnetic charge and $\alpha > 0$ is the GB coupling constant. Again we value the usual parameter, $m=1$. If we solve the metric function (20) based on the assumed values to find the event horizon in terms of $\alpha = 0.4$(optional), we will find that our metric function has a solution for $ -0.7026\leq g \leq 0.70259 $, fig(6), and for values outside this range, the metric function is always unanswered, which means that it will be in the form of a naked singularity.
\begin{figure}[h!]
 \begin{center}
 \includegraphics[height=7.5cm,width=6.8cm]{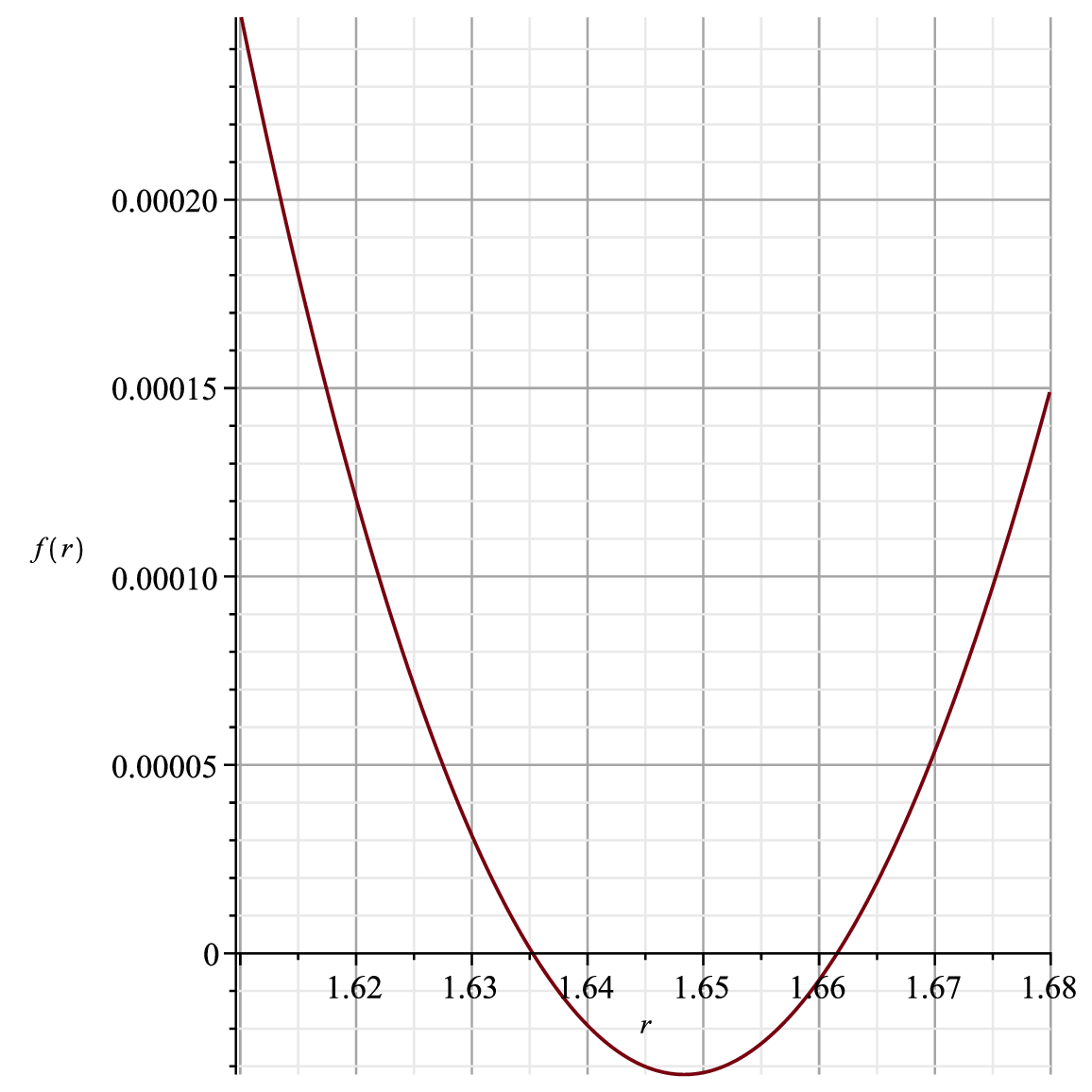}
 \caption{\small{this plot represents ($f(r)$ vs. r) with $\alpha = 0.4$ and $g = 0.7026 $ where the event horizon located at $ r_+ = 1.661555644$}}
 \label{2}
\end{center}
\end{figure}
According to the parametric values mentioned above and with respect to (21) and (22), for such black hole, the potential H and $\phi$ (15),(16) are,
\begin{equation}\label{31}
H =\frac{\sqrt{4+\frac{2 r^{2} \left(1-\sqrt{1+\frac{8 \alpha}{\left(g^{2}+r^{2}\right)^{\frac{4}{3}}}}\right)}{\alpha}}}{2 \sin \! \left(\theta \right) r},
\end{equation}
\begin{equation}\label{32}
\phi^{r}=-\frac{\left(g^{2}+r^{2}\right)^{\frac{7}{3}} \sqrt{\frac{\left(g^{2}+r^{2}\right)^{\frac{4}{3}}+8 \alpha}{\left(g^{2}+r^{2}\right)^{\frac{4}{3}}}}-\frac{8 r^{4}}{3}}{\sqrt{\frac{\left(g^{2}+r^{2}\right)^{\frac{4}{3}}+8 \alpha}{\left(g^{2}+r^{2}\right)^{\frac{4}{3}}}}\, \left(g^{2}+r^{2}\right)^{\frac{7}{3}} r^{2} \sin \! \left(\theta \right)},
\end{equation}
\begin{equation}\label{33}
\phi^{\theta}=-\frac{\sqrt{4+\frac{2 r^{2} \left(1-\sqrt{1+\frac{8 \alpha}{\left(g^{2}+r^{2}\right)^{\frac{4}{3}}}}\right)}{\alpha}}\, \cos \! \left(\theta \right)}{2 \sin \! \left(\theta \right)^{2} r^{2}}.
\end{equation}
Calculations for this structure show that, the black hole in the allowed range $ -0.7026\leq g \leq 0.70259 $ in fig (7a) shows a total topological charge of -1 and has an unstable photon sphere. In this fig again two topological charges exist and therefore the total charge must be zero.
but the event horizon located at $r_+ = 1.661555644$, that is, one of the charge is located behind the horizon and practically according to the concept of WCC and it will not be included in the calculations. Calculations show that outside of the above normal range, the structure will remain naked singularity form only up to g=0.865, that is, the total topological charge will be zero, and after this value, the solution will be practically lost. The full results are in Table 3.
\begin{figure}[h!]
 \begin{center}
 \subfigure[]{
 \includegraphics[height=5.5cm,width=6cm]{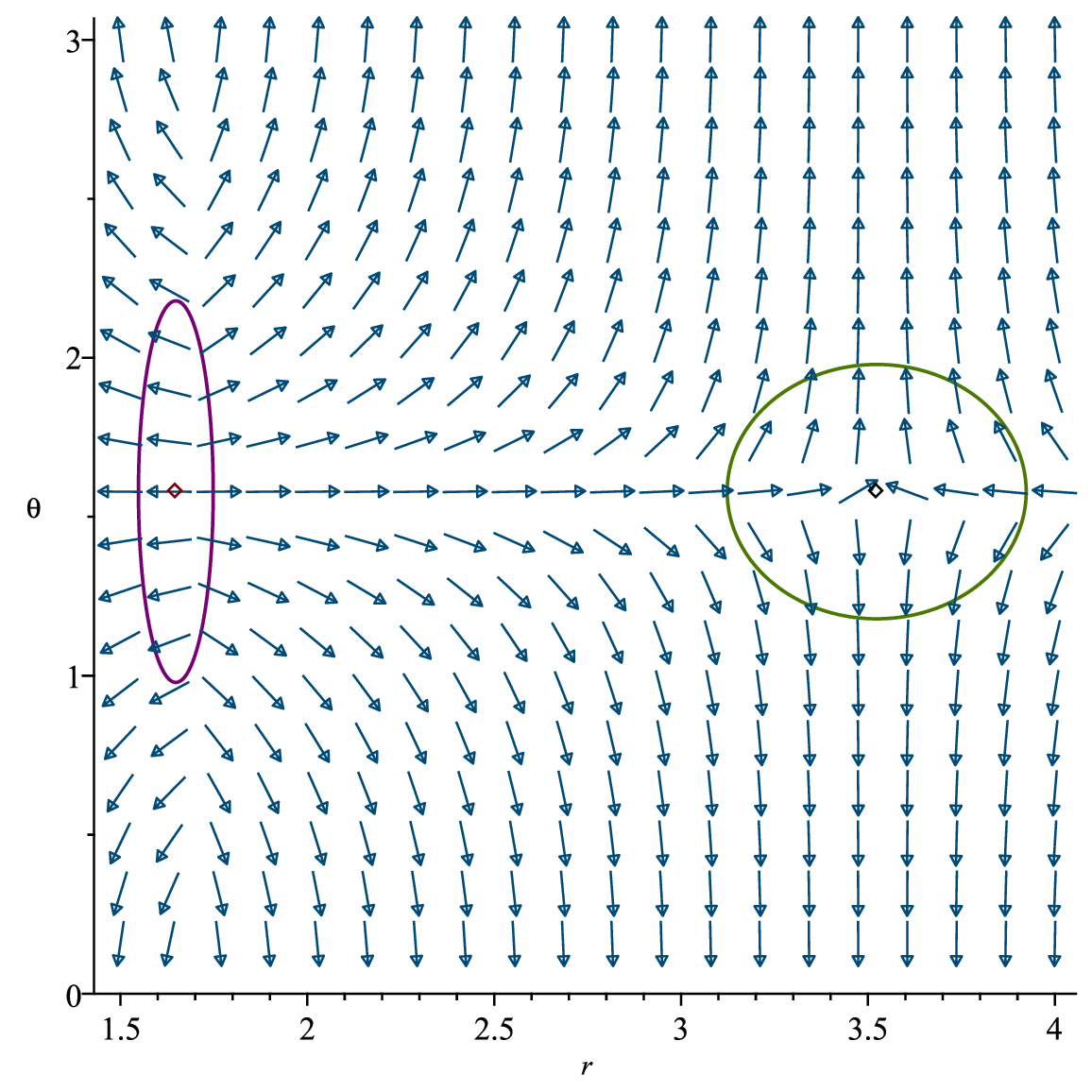}
 \label{7a}}
 \subfigure[]{
 \includegraphics[height=5.5cm,width=6cm]{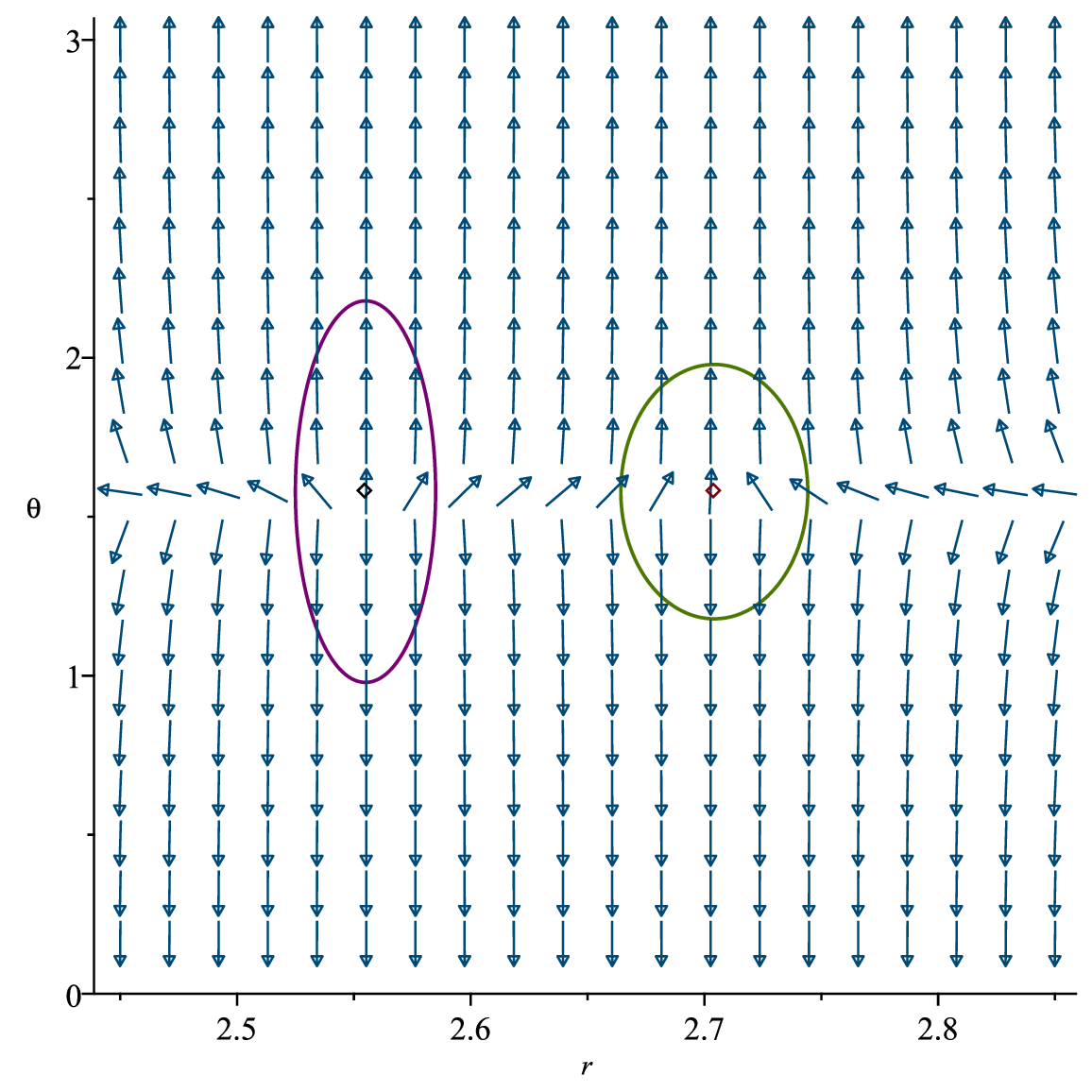}
 \label{7b}}
   \caption{\small{Plot 7(a) the photon spheres is located at $ (r,\theta)=(1.648235913,1.57)$ and $ (r,\theta)=(3.523343153,1.57)$ with respect to $(g=0.70259,m=1,\alpha = 0.4)$ , plot 7(b) the photon spheres is located at $ (r,\theta)=(2.554954982,1.57)$ and $ (r,\theta)=(2.704224286,1.57)$ with respect to $(g=0.865,m=1,\alpha = 0.4)$ }}
 \label{5}
\end{center}
 \end{figure}
\begin{center}
\begin{table}[h!]
  \centering
\begin{tabular}{|p{3.2cm}|p{4cm}|p{4.5cm}|p{1.5cm}|p{2cm}|}
  \hline
  \centering{5D Bardeen black holes in EGB}  & \centering{Fix parameters} &\centering{Conditions}&*TTC&\ $(R_{PLPS})$\\[3mm]
   \hline
  \centering{unstable photon sphere} & $m=1,\alpha = 0.4$ & \centering{$-0.7026\leq g \leq 0.70259$} &\centering{ $-1$}&\ $3.523343153$\\[2.5mm]
   \hline
  \centering{naked singularity} & $m=1,\alpha = 0.4$ & \centering{$-0.865\leq g \leq -0.70259$\\ $0.70259 \leq g\leq 0.865$} &\centering{ $0$}&\ $-$ \\[4.5mm]
   \hline
   \centering{*Unauthorized area} & $m=1,\alpha = 0.4$ & \centering{$g<-0.865$\\$g>0.865 $} & $ nothing$&\ $-$\\[2.5mm]
   \hline
      \end{tabular}
   \caption{*Unauthorized region: is the region where the roots of $\phi$ equations become negative or imaginary in this region.\\ TTC: *Total Topological Charge\\}\label{1}
\end{table}
\end{center}
\subsection{$4D:\Lambda\neq0 $}
Since the behavior is very similar to the 5D model, we only mention the main formulas and show the results in the final table4. \cite{54.0}
\begin{equation}\label{(34)}
f =1+\frac{r^{2} \left(1-\sqrt{1+4 \left(\frac{2 m}{\left(g^{2}+r^{2}\right)^{\frac{3}{2}}}-\frac{1}{l^{2}}\right) \alpha}\right)}{2 \alpha},
\end{equation}
\begin{equation}\label{(35)}
H =\frac{\sqrt{4+\frac{2 r^{2} \left(1-\sqrt{1+4 \left(\frac{2}{\left(g^{2}+r^{2}\right)^{\frac{3}{2}}}-1\right) \alpha}\right)}{\alpha}}}{2 \sin \! \left(\theta \right) r},
\end{equation}
\begin{equation}\label{(36)}
\phi^{r}=-\frac{2 \left(g^{2}+r^{2}\right)^{\frac{5}{2}} \sqrt{-\frac{\left(\alpha -\frac{1}{4}\right) \left(g^{2}+r^{2}\right)^{\frac{3}{2}}-2 \alpha}{\left(g^{2}+r^{2}\right)^{\frac{3}{2}}}}-3 r^{4}}{2 \sqrt{-\frac{\left(\alpha -\frac{1}{4}\right) \left(g^{2}+r^{2}\right)^{\frac{3}{2}}-2 \alpha}{\left(g^{2}+r^{2}\right)^{\frac{3}{2}}}}\, \left(g^{2}+r^{2}\right)^{\frac{5}{2}} r^{2} \sin \! \left(\theta \right)},
\end{equation}
\begin{equation}\label{(37)}
\phi^{\theta}=-\frac{\sqrt{4+\frac{2 r^{2} \left(1-\sqrt{1+4 \left(\frac{2}{\left(g^{2}+r^{2}\right)^{\frac{3}{2}}}-1\right) \alpha}\right)}{\alpha}}\, \cos \! \left(\theta \right)}{2 \sin \! \left(\theta \right)^{2} r^{2}}.
\end{equation}
\begin{center}
\begin{table}[h!]
  \centering
\begin{tabular}{|p{3.2cm}|p{4cm}|p{4.5cm}|p{1.5cm}|p{2cm}|}
  \hline
  \centering{4D Bardeen AdS black holes in EGB}  & \centering{Fix parameters} &\centering{Conditions}&*TTC&\ $(R_{PLPS})$\\[3mm]
     \hline
  \centering{unstable photon sphere} & $m=1,\alpha = 0.4,l=1$ & \centering{$-0.401\leq g \leq 0.401$} &\centering{ $-1$}&\ $6.553871469$\\[2.5mm]
   \hline
  \centering{naked singularity} & $m=1,\alpha = 0.4,l=1$ & \centering{$-1.824\leq g \leq -0.401$\\ $0.401 \leq g\leq 1.824$} &\centering{ $0$}&\ $-$ \\[4.5mm]
   \hline
   \centering{*Unauthorized area} & $m=1,\alpha = 0.4,l=1$ & \centering{$g<-1.824$\\$g>1.824 $} & $ nothing$&\ $-$\\[2.5mm]
   \hline
      \end{tabular}
   \caption{*Unauthorized region: is the region where the roots of $\phi$ equations become negative or imaginary in this region.\\ TTC: *Total Topological Charge\\}\label{1}
\end{table}
\end{center}
\section{Conclusions}
This article can be analyzed from two perspectives. In the first and simplest view, we investigate the existence and behavioral form of photon spheres and anti-photon spheres, based on the total topological charges \cite{37}, for black holes with different structures, such as non-minimal Einstein-Yang-Mills, Dark Fluid AdS Chaplygin-like, and Bardeen black holes within regular Einstein-Gauss-Bonnet (EGB) gravity. The graphs of the fields and the zero points are presented in Figures (2), (3), (5), and (7), and the properties of the structure under study are detailed in Tables (1-4). These models, which might have posed more complex computational challenges if we had pursued conventional methods, namely deriving the Hamiltonian from the action and then calculating the effective potential for the photon sphere, are less problematic when approached through topological methods.\\
From a deeper perspective, if we want to  discuss our primary motivation for these calculations, it is essential to acknowledge that the existence of a photon sphere for ultra-compact gravitational structures like black holes is a necessity which is largely confirmed by studies conducted on various black hole models\cite{20,21,22,23,24,25,26,27,28,29,30,31,55,56,57,58,59,60,61,62,63,64,65,66,67,68}.\\
Leveraging this necessity can bring several interesting implications. As the simplest outcome, the calculation of unstable photon spheres could obviate the need for a simple computation of the WCCC, as an unstable photon sphere will only dominate the space of the system when an event horizon exists, and thus the WCCC condition is satisfied.\\
However, the most significant implication, which shapes the main goal of our paper, is that we can simply classify the parameter space of the models under study based on the existence and location of the topological photon spheres and anti-photon spheres. That is, we can accurately demonstrate which parameter ranges of the model correspond to a black hole structure, a naked singularity, or even a model that, mathematically, lacks a physical and scientific response for its surrounding space, which we have termed the forbidden region.\\
The other achievement of the necessity of the photon sphere existing and its integration with a topological perspective is that we can also use the aforementioned classification inversely. In a process akin to reverse engineering, we can balance the parameters of our system in such a way that the model under study is compelled to exhibit the behavior we desire. This outcome is quite intriguing in its own right, as in certain gravitational studies, we require a gravitational structure that adheres to specific principles and behaviors. In such cases, this method could serve as a significant contribution to these studies, which necessitate particular parameter ranges while simultaneously requiring the structure to remain in the form of a black hole or any other format. To elucidate this point further, let us provide an example. In the study model of the Swampland Conjecture, there are hypotheses such as the Weak Gravity Conjecture (WGC), among others, which require that the black hole, while becoming overcharged, must still retain a quasi-black hole form, or for instance, that the strength of terms related to magnetic parameters be compared and influenced by electric and gravitational counterparts.
In this work, we suggest that for such scenarios, one can utilize the topological behavior of the photon sphere and, after adjusting the parameters to the desired form, examine the topological photon sphere to determine whether the desired conditions can be achieved.\\
In summary, and as a synthesis of all the above discussions, it can be proposed that a bidirectional relationship exists between the parameters and the topological photon sphere. On one hand, based on the behavior of the topological photon sphere, the permissible range for the parameters of the model under study can be determined, and the space surrounding the model can be accurately classified according to the behavior of the photon sphere. Conversely, by choosing the parameters in the form required in the study, they can be used as a tool to control the behavior of the black hole.\\ 
Finally, we remark that in this article we focus on the specific parameters of each model, such as the non-minimal coupling, the Chaplygin gas parameter, and the Bardeen charge. However, if these parameters are fixed, the method of range finding can be extended to the common and main parameters of the black holes, such as the mass, the charge, and the angular momentum. We leave this generalization for future work.


\newpage
\begin{thebibliography}{11}
\bibitem{1}
Goswami, Rituparno, Pankaj S. Joshi, and Parampreet Singh. "Quantum evaporation of a naked singularity." Physical review letters 96.3 (2006): 031302.
\bibitem{2}
Goswami, Rituparno, and Pankaj S. Joshi. "Spherical gravitational collapse in N dimensions." Physical Review D 76.8 (2007): 084026.
\bibitem{3}
Christodoulou, Demetrios. "The instability of naked singularities in the gravitational collapse of a scalar field." Annals of Mathematics 149.1 (1999): 183-217.
\bibitem{4}
Pugliese, Daniela, and Hernando Quevedo. "Naked singularities and black hole Killing horizons." arXiv preprint arXiv:2402.07512 (2024).
\bibitem{5}
Viththani, Divyesh P., et al. "Particle motion and tidal force in a non-vacuum-charged naked singularity." arXiv preprint arXiv:2402.02069 (2024).
\bibitem{6}
Joshi, Pankaj S., and Sudip Bhattacharyya. "Primordial naked singularities." arXiv preprint arXiv:2401.14431 (2024).
\bibitem{7}
Deliyski, Valentin, et al. "Observing naked singularities by the present and next-generation Event Horizon Telescope." arXiv preprint arXiv:2401.14092 (2024).
\bibitem{8}
Ghosh, Avisikta. "Study of Orbital Dynamics in Singular and Regular Naked Singularity Space-times." arXiv preprint arXiv:2401.10771 (2024).
\bibitem{9}
Chen, Yiqian, Peng Wang, and Haitang Yang. "Observations of Orbiting Hot Spots around Naked Singularities." arXiv preprint arXiv:2309.04157 (2023).
\bibitem{10}
Wang, Mingzhi, et al. "The images of a rotating naked singularity with a complete photon sphere." arXiv preprint arXiv:2307.16748 (2023).
\bibitem{11}
Joshi, P. S., and I. H. Dwivedi. "Naked singularities in spherically symmetric inhomogeneous Tolman-Bondi dust cloud collapse." Physical Review D 47.12 (1993): 5357.
\bibitem{12}
Goswami, Rituparno, and Pankaj S. Joshi. "Gravitational collapse of an isentropic perfect fluid with a linear equation of state." Classical and Quantum Gravity 21.15 (2004): 3645.
\bibitem{13}
Mosani, Karim, Dipanjan Dey, and Pankaj S. Joshi. "Strong curvature naked singularities in spherically symmetric perfect fluid collapse." Physical Review D 101.4 (2020): 044052.
\bibitem{14}
Acharya, Kauntey, et al. "Naked Singularity as a Possible Source of Ultra-High Energy Cosmic Rays." arXiv preprint arXiv:2303.16590 (2023).
\bibitem{15}
Mosani, Karim, Dipanjan Dey, and Pankaj S. Joshi. "Globally visible singularity in an astrophysical setup." Monthly Notices of the Royal Astronomical Society 504.4 (2021): 4743-4750.
\bibitem{16}
Dey, Dipanjan, et al. "Causal structure of singularity in non-spherical gravitational collapse." The European Physical Journal C 82.5 (2022): 431.
\bibitem{17}
Deshingkar, S. S., S. Jhingan, and P. S. Joshi. "On the Global Visibility of the Singularity in Quasi-Spherical Collapse." General Relativity and Gravitation 30.10 (1998): 1477-1499.
\bibitem{18}
Acharya, Kauntey, et al. "Naked Singularity as a Possible Source of Ultra-High Energy Cosmic Rays." arXiv preprint arXiv:2303.16590 (2023).
\bibitem{19}
Joshi, Pankaj S., Daniele Malafarina, and Ramesh Narayan. "Equilibrium configurations from gravitational collapse." Classical and Quantum Gravity 28.23 (2011): 235018.
\bibitem{20}
Event Horizon Telescope Collaboration. "First M87 event horizon telescope results. IV. Imaging the central supermassive black hole." arXiv preprint arXiv:1906.11241 (2019).
\bibitem{21}
Akiyama, Kazunori, et al. "First M87 event horizon telescope results. VI. The shadow and mass of the central black hole." The Astrophysical Journal Letters 875.1 (2019): L6.
\bibitem{22}
Abbott, Benjamin P., et al. "Observation of gravitational waves from a binary black hole merger." Physical review letters 116.6 (2016): 061102.
\bibitem{23}
Abbott, B. P., et al. "GWTC-1: a gravitational-wave transient catalog of compact binary mergers observed by LIGO and Virgo during the first and second observing runs." Physical Review X 9.3 (2019): 031040.
\bibitem{24}
Akiyama, Kazunori, et al. "First M87 event horizon telescope results. VI. The shadow and mass of the central black hole." The Astrophysical Journal Letters 875.1 (2019): L6. 041301.
\bibitem{25}
Akiyama, Kazunori, et al. "First M87 event horizon telescope results. V. Physical origin of the asymmetric ring." The Astrophysical Journal Letters 875.1 (2019): L5.
\bibitem{26}
Riojas, Marcos, and Hao-Yu Sun. "The Photon Sphere and the AdS/CFT Correspondence." arXiv preprint arXiv:2307.06415 (2023).
\bibitem{27}
Hashimoto, Koji, et al. "Photon sphere and quasinormal modes in AdS/CFT." Journal of High Energy Physics 2023.10 (2023): 1-39.
\bibitem{28}
Liu, Yuxuan, et al. "Holographic Einstein ring of a charged AdS black hole." Journal of High Energy Physics 2022.10 (2022): 1-19.
\bibitem{29}
Wang, Mingzhi, et al. "The images of a rotating naked singularity with a complete photon sphere." arXiv preprint arXiv:2307.16748 (2023).
\bibitem{30}
Guo, Guangzhou, et al. "Black holes with multiple photon spheres." Physical Review D 107.12 (2023): 124037.
\bibitem{31}
Ladino, Jose Miguel, and Eduard Larrañaga. "Eikonal quasinormal modes, photon sphere and shadow of a charged black hole in the 4D Einstein-Gauss-Bonnet gravity." arXiv preprint arXiv:2303.02877 (2023).
\bibitem{32}
Cardoso, Vitor, et al. "Light rings as observational evidence for event horizons: long-lived modes, ergoregions and nonlinear instabilities of ultracompact objects." Physical Review D 90.4 (2014): 044069.
\bibitem{33}
Cunha, Pedro VP, Carlos AR Herdeiro, and Eugen Radu. "Fundamental photon orbits: black hole shadows and spacetime instabilities." Physical Review D 96.2 (2017): 024039.
\bibitem{34}
Cunha, Pedro VP, Emanuele Berti, and Carlos AR Herdeiro. "Light-ring stability for ultracompact objects." Physical review letters 119.25 (2017): 251102.
\bibitem{35}
Ghosh, Rajes, and Sudipta Sarkar. "Light rings of stationary spacetimes." Physical Review D 104.4 (2021): 044019.
\bibitem{36}
Cunha, Pedro VP, and Carlos AR Herdeiro. "Stationary black holes and light rings." Physical Review Letters 124.18 (2020): 181101.
\bibitem{37}
Wei, Shao-Wen. "Topological charge and black hole photon spheres." Physical Review D 102.6 (2020): 064039.
\bibitem{38}
Duan, Y. S. "The structure of the topological current." Preprint SLAC-PUB-3301/84 (1984).
\bibitem{39}
Sadeghi, Jafar, et al. "Thermodynamic topology and photon spheres in the hyperscaling violating black holes." Astroparticle Physics 156 (2024): 102920.
\bibitem{40}
Balakin, Alexander B., Jose PS Lemos, and Alexei E. Zayats. "Regular nonminimal magnetic black holes in spacetimes with a cosmological constant." Physical Review D 93.2 (2016): 024008.
\bibitem{41}
Balakin, A. B., H. Dehnen, and A. E. Zayats. "Nonminimal Einstein-Yang-Mills-Higgs theory: Associated, color, and color-acoustic metrics for the Wu-Yang monopole model." Physical Review D 76.12 (2007): 124011.
\bibitem{42}
Balakin, Alexander B., and Alexei E. Zayats. "Non-minimal Wu–Yang monopole." Physics Letters B 644.5-6 (2007): 294-298. (2023).
\bibitem{43}
Kala, Shubham, Hemwati Nandan, and Prateek Sharma. "Shadow and weak gravitational lensing of a rotating regular black hole in a non-minimally coupled Einstein-Yang-Mills theory in the presence of plasma." The European Physical Journal Plus 137.4 (2022): 1-18.
\bibitem{44}
Al-Badawi, Ahmad, and M. Q. Owaidat. "Particle dynamics and shadow of a regular non-minimal magnetic black hole." General Relativity and Gravitation 55.11 (2023): 131.
\bibitem{45}
Gogoi, Dhruba Jyoti, and Supakchai Ponglertsakul. "Constraints on Quasinormal modes from Black Hole Shadows in regular non-minimal Einstein Yang-Mills Gravity." arXiv preprint arXiv:2402.06186 (2024).
\bibitem{46}
Zhang, Ruanjing, et al. "Strong gravitational lensing of rotating regular black holes in non-minimally coupled Einstein-Yang-Mills theory." arXiv preprint arXiv:2304.13263
\bibitem{47}
Li, Xiang-Qian, et al. "Critical behavior of AdS black holes surrounded by dark fluid with Chaplygin-like equation of state." Physical Review D 107.10 (2023): 104055.
\bibitem{48}
Sekhmani, Y., et al. "Phase structure of charged AdS black holes surrounded by exotic fluid with modified Chaplygin equation of state." arXiv preprint arXiv:2311.02448 (2023).
\bibitem{49}
Zhang, Meng-Yao, et al. "Critical behavior and Joule-Thomson expansion of charged AdS black holes surrounded by exotic fluid with modified Chaplygin equation of state." arXiv preprint arXiv:2401.17589 (2024).
\bibitem{50}
Singh, Dharm Veer, Sushant G. Ghosh, and Sunil D. Maharaj. "Bardeen-like regular black holes in 5D Einstein–Gauss–Bonnet gravity." Annals of Physics 412 (2020): 168025.
\bibitem{51}
Kumar, Arun, Rahul Kumar Walia, and Sushant G. Ghosh. "Bardeen Black Holes in the Regularized 4 D Einstein–Gauss–Bonnet Gravity." Universe 8.4 (2022): 232.
\bibitem{52}
Kumar, Amit, et al. "Exact solution of Bardeen black hole in Einstein–Gauss–Bonnet gravity." The European Physical Journal Plus 138.12 (2023): 1-14.
\bibitem{53}
Islam, Shafqat Ul, Sushant G. Ghosh, and Sunil D. Maharaj. "Strong gravitational lensing by Bardeen black holes in 4D EGB gravity: constraints from supermassive black holes." arXiv preprint arXiv:2203.00957 (2022).
\bibitem{54.0}
Singh, Dharm Veer, and Sanjay Siwach. "Thermodynamics and Pv criticality of Bardeen-AdS black hole in 4D Einstein-Gauss-Bonnet gravity." Physics Letters B 808 (2020): 135658.
\bibitem{55}
Koga, Yasutaka, and Tomohiro Harada. "Stability of null orbits on photon spheres and photon surfaces." Physical Review D 100.6 (2019): 064040.
\bibitem{56}
da Silva, Luís F. Dias, et al. "Photon rings as tests for alternative spherically symmetric geometries with thin accretion disks." Physical Review D 108.8 (2023): 084055.
\bibitem{57}
Hod, Shahar. "On the number of light rings in curved spacetimes of ultra-compact objects." Physics Letters B 776 (2018): 1-4.
\bibitem{58}
Bianchi, Massimo, et al. "Light rings of five-dimensional geometries." Journal of High Energy Physics 2021.3 (2021): 1-24.
\bibitem{59}
Grandclément, Philippe. "Light rings and light points of boson stars." Physical Review D 95.8 (2017): 084011.
\bibitem{60}
Koga, Yasutaka, et al. "Dynamical photon sphere and time evolving shadow around black holes with temporal accretion." Physical Review D 105.10 (2022): 104040.
\bibitem{61}
Amir, Muhammed, Balendra Pratap Singh, and Sushant G. Ghosh. "Shadows of rotating five-dimensional charged EMCS black holes." The European Physical Journal C 78 (2018): 1-15.
\bibitem{62}
Franzin, Edgardo, Stefano Liberati, and Vania Vellucci. "From regular black holes to horizonless objects: quasi-normal modes, instabilities and spectroscopy." Journal of Cosmology and Astroparticle Physics 2024.01 (2024): 020.
\bibitem{63}
Heydarzade, Yaghoub, and Vitalii Vertogradov. "Dynamical Photon Spheres in Charged Black Holes and Naked Singularities." arXiv e-prints (2023): arXiv-2311.
\bibitem{64}
Zubair, M., Muhammad Ali Raza, and Ghulam Abbas. "Optical features of rotating black hole with nonlinear electrodynamics." The European Physical Journal C 82.10 (2022): 948.
\bibitem{65}
Solanki, Jay, and Volker Perlick. "Photon sphere and shadow of a time-dependent black hole described by a Vaidya metric." Physical Review D 105.6 (2022): 064056.
\bibitem{66}
Guo, Sen, et al. "The shadow and photon sphere of the charged black hole in Rastall gravity." Classical and Quantum Gravity 38.16 (2021): 165013.
\bibitem{67}
Cunha, Pedro VP, Carlos AR Herdeiro, and João PA Novo. "Light rings on stationary axisymmetric spacetimes: blind to the topology and able to coexist." Physical Review D 109.6 (2024): 064050.
\bibitem{68}
Wei, Shao-Wen, et al. "Static spheres around spherically symmetric black hole spacetime." Physical Review Research 5.4 (2023): 043050.

\end{thebibliography}
\end{document}